\documentclass[11pt]{article}
\usepackage{amsmath,amssymb,color,epsfig,cite}
\usepackage{graphicx}
\usepackage{subfigure}
\usepackage{setspace}
\usepackage{amsthm,mathenv}

\allowdisplaybreaks[4]

\usepackage{amsfonts}

\newcommand{\hoch}[1]{$\, ^{#1}$}

\makeatletter
\@addtoreset{equation}{section}
\makeatother

\newcommand{\be}{\begin{equation}}
\newcommand{\ee}{\end{equation}}
\newcommand{\bea}{\setlength\arraycolsep{2pt} \begin{eqnarray}}
\newcommand{\eea}{\end{eqnarray}}
\newcommand{\nn}{\nonumber}

\def\0{{\sst{(0)}}}
\def\1{{\sst{(1)}}}
\def\2{{\sst{(2)}}}
\def\3{{\sst{(3)}}}
\def\4{{\sst{(4)}}}
\def\5{{\sst{(5)}}}
\def\6{{\sst{(6)}}}
\def\7{{\sst{(7)}}}
\def\8{{\sst{(8)}}}
\def\sst#1{{\scriptscriptstyle #1}}

\begin{document}

\begin{center}
{\large {\bf NANOGrav signal from first-order confinement/deconfinement phase transition in different QCD matters}}

\vspace{15pt}
{\large Shou-Long Li\hoch{1}, Lijing Shao\hoch{2, 3}, Puxun Wu\hoch{1} and Hongwei Yu\hoch{1}}

\vspace{10pt}

\hoch{1}{\it Department of Physics and Synergetic Innovation Center for Quantum Effect and Applications, Hunan Normal University, Changsha 410081, China}

\hoch{2}{\it Kavli Institute for Astronomy and Astrophysics, Peking University, Beijing 100871, China}

\hoch{3}{\it National Astronomical Observatories, Chinese Academy of Sciences, Beijing 100012, China}

\vspace{40pt}

\underline{ABSTRACT}

\end{center}

Recently, an indicative evidence of a stochastic process, reported by the
NANOGrav Collaboration based on the analysis of 12.5-year pulsar timing
array data which might be interpreted as a potential stochastic
gravitational wave signal, has aroused keen interest of theorists. The
first-order color charge confinement phase transition at the QCD
scale could be one of the cosmological sources for the NANOGrav signal. If
the phase transition is flavor dependent and happens sequentially, it is
important to find that what kind of QCD matter in which the first-order confinement/deconfinement phase transition happens is more likely to be the potential source of the NANOGrav signal during the evolution of the universe.
 In this paper, we would like to illustrate that the NANOGrav signal could
be generated from confinement/deconfinement transition in either heavy
static quarks with a zero baryon chemical potential, or quarks with a
finite baryon chemical potential.  In contrast, the gluon
confinement could not possibly be  the source for the NANOGrav signal according to the current observation. 
Future observation will help to distinguish between different
scenarios.

\vfill
 shoulongli@hunnu.edu.cn\ \ \ lshao@pku.edu.cn \ \ \ pxwu@hunnu.edu.cn \ \ \ hwyu@hunnu.edu.cn

\thispagestyle{empty}

\pagebreak

\newpage

\section{Introduction}

Recently, the North American Nanohertz Observatory for Gravitational Wave
(NANOGrav) Collaboration~\cite{Arzoumanian:2020vkk} has reported an
analysis of 12.5-year pulsar timing array (PTA) data. According to the
analysis~\cite{Arzoumanian:2020vkk}, a possible evidence is found for a
stochastic common-spectrum process which may be interpreted as a
gravitational wave (GW) signal with its frequency in $1$--$10$ nHz, and its
average GW energy density $\langle \Omega_{\textup{GW}} h^2
\rangle_{\textup{NANOGrav}} \sim 10^{-10}$ with an almost flat GW spectrum
$\Omega_{\rm GW} h^2 \sim f^{-1.5\pm 0.5 } $ at 1-$\sigma$ level. Although
the observational results need further analyses, such as a joint analysis
with data from the other PTA collaborations (such as EPTA and
PPTA)~\cite{Desvignes:2016yex, Kerr:2020qdo, Perera:2019sca}, the potentiality of its being a
stochastic GW background (SGWB) signal has aroused keen interest of
theorists~\cite{Ellis:2020ena, Blasi:2020mfx, Vaskonen:2020lbd,
DeLuca:2020agl, Buchmuller:2020lbh, Nakai:2020oit, Addazi:2020zcj,
Ratzinger:2020koh, Kohri:2020qqd, Samanta:2020cdk, Vagnozzi:2020gtf,
Bian:2021lmz, Neronov:2020qrl, Li:2020cjj, Paul:2020wbz, Domenech:2020ers,
Bhattacharya:2020lhc, Abe:2020sqb, Inomata:2020xad, Middleton:2020asl,
Kuroyanagi:2020sfw, Pandey:2020gjy, Ramberg:2020oct, Barman:2020jrf,
Atal:2020yic, Chen:2021wdo, Bigazzi:2020avc,Datta:2020bht}. On the one hand, one may explain it as a potential
SGWB signal by considering different astrophysical and cosmological
sources. On the other hand, the possible SGWB signal could serve as a potential new
probe to studying new physics.

During the evolution of the universe, as the temperature decreases, the
universe may undergo several phase transitions from the metastable vacuums to
stable vacuums.  At the quantum chromodynamics (QCD) energy scale, there are
two important transitions, i.e. the spontaneous chiral symmetry breaking and
the color charge confinement. For the confinement transition, the universe
will go from a quark-gluon plasma (QGP) phase to a hadron phase as
the temperature decreases. The numerical lattice simulation
shows~\cite{Aoki:2006br, Bazavov:2011nk, Bhattacharya:2014ara} that the QCD
transition is likely a crossover for three dynamical quark flavors when the
baryon and charge chemical potential is negligible, i.e., in the absence of
the baryon and lepton asymmetries.  If the transition is first-order, GWs could be produced due to the violent process of vacuum bubble
nucleation and subsequent bubble collisions~\cite{Kosowsky:1991ua,
Caprini:2015zlo, Kosowsky:1992rz, Kosowsky:1992vn, Kamionkowski:1993fg,
Caprini:2007xq, Huber:2008hg}, sound waves~\cite{Hindmarsh:2013xza,
Giblin:2013kea, Giblin:2014qia, Hindmarsh:2015qta} and magnetohydrodynamic
(MHD) turbulence~\cite{Kosowsky:2001xp, Caprini:2006jb,
Kahniashvili:2008pf, Kahniashvili:2008pe, Kahniashvili:2009mf,
Caprini:2009yp, Kisslinger:2015hua}. The GWs produced are within the
frequency range of the PTA observation~\cite{Caprini:2010xv, Anand:2017kar,
Ahmadvand:2017xrw}. The three processes in the first-order phase
transition could be potential sources of the NANOGrav
signal~\cite{Paul:2020wbz, Neronov:2020qrl, Abe:2020sqb}. Therefore,  
the next issue is what are the QCD matters in which the first-order phase transition 
can occur. In this regard, let us note that  for the case of a zero baryon chemical potential, the
transition is first-order in a non-dynamical (static) heavy quark
system~\cite{Philipsen:2010gj, Petreczky:2012rq}.  For the case of a finite
baryon chemical potential,  a large lepton asymmetry
might affect the dynamics of the QCD phase transition in a way to render it first-order in the early universe~\cite{Schwarz:2009ii}.
 Besides, for
a pure gluon system, the first-order phase transition might occur as well.

Different QCD matter has a different temperature of phase transition which is a crucial parameter determining the features of  the GWs produced. Then a question arises naturally 
as to  what kind of
QCD matter in which the first-order confinement/deconfinement phase transition
happens is more likely to  be the potential source of the NANOGrav signal
during the evolution of the universe. 
 In this paper, we will try to answer
this question. We consider three types of QCD matter systems: (i) heavy
static quarks with a zero baryon chemical potential, (ii) quarks with a
finite baryon chemical potential, and (iii) a pure gluon system. We match
the NANOGrav signal with the GW spectra  from the first-order confinement/deconfinement phase transitions in three QCD matter systems with holographic
models~\cite{Ahmadvand:2017xrw, Ahmadvand:2017tue, Chen:2017cyc}. We show
that the GW spectra from the phase transitions in pure quark systems, regardless of whether the chemical potential is finite or zero, could 
explain the NANOGrav signal according to the current observation. In contrast, the signal could not possibly come from the
 gluon confinement.

\section{GWs from  first-order phase transitions}

When first-order phase transition occurs, the universe transfers from a
metastable vacuum to a stable vacuum. This process can be described as
bubble nucleation. Generally, the GW signal from cosmological first-order
phase transitions mainly comes from three processes: collisions of vacuum
bubble walls, sound waves, and the MHD turbulence in the
plasma~\cite{Kosowsky:1991ua, Caprini:2015zlo, Kosowsky:1992rz,
Kosowsky:1992vn, Kamionkowski:1993fg, Caprini:2007xq, Huber:2008hg,
Hindmarsh:2013xza, Giblin:2013kea, Giblin:2014qia, Hindmarsh:2015qta,
Kosowsky:2001xp, Caprini:2006jb, Kahniashvili:2008pf, Kahniashvili:2008pe,
Kahniashvili:2009mf, Caprini:2009yp, Kisslinger:2015hua}. 
So, the total energy
density of the GW, which is a sum of  the three, is given by
\be
h^2 \Omega (f) \simeq h^2 \Omega_{\textup{en}} (f) + h^2 \Omega_{\textup{sw}} (f) + h^2 \Omega_{\textup{tu}} (f) \,, \label{GW}
\ee
where $h = H_0/100 \ \textup{km}^{-1}\ \textup{s}\ \textup{Mpc} $ in which $H_0$ is the Hubble constant today.  $h^2 \Omega_{\textup{en}}$, $h^2 \Omega_{\textup{sw}}$ and $h^2
\Omega_{\textup{tu}}$ are the contributions from bubble collision, sound
waves, and the turbulence, respectively, which are given by~\cite{Huber:2008hg,
Caprini:2015zlo, Hindmarsh:2015qta, Caprini:2009yp}
\bea
h^2 \Omega_{\textup{en}}  &=& 3.6 \times 10^{-5} \left(\frac{H_\ast}{\beta} \right)^2 \left(\frac{\kappa_1 \alpha}{1+\alpha} \right)^2 \left(\frac{10}{g_\ast} \right)^{\frac13} \left(\frac{0.11 v_w^3}{0.42+v_w^2} \right) S_{\textup{en}} \,,\nn \\
h^2 \Omega_{\textup{sw}} &=& 5.7 \times 10^{-6} \left(\frac{H_\ast}{\beta} \right) \left(\frac{\kappa_2 \alpha}{1+\alpha}\right)^2 \left(\frac{10}{g_\ast} \right)^{\frac13}  v_w S_{\textup{sw}}  \,, \nn \\
h^2 \Omega_{\textup{tu}}  &=& 7.2 \times 10^{-4} \left(\frac{H_\ast}{\beta} \right) \left(\frac{\kappa_3 \alpha}{1+\alpha} \right)^2 \left(\frac{10}{g_\ast} \right)^{\frac13} v_w S_{\textup{tu}}  \,. \label{eq:Omega}
\eea
In above equations, $ \beta$ is the inverse time duration of the phase
transition, $v_w$ is the velocity of bubble wall, $g_\ast $ and $ H_\ast$
represent the number of active degrees of freedom and Hubble parameter at
the time of production of GWs respectively, $\alpha = 30 \epsilon_\ast /
(\pi^2 g_\ast T_\ast^4) $ is the ratio of the vacuum energy density
$\epsilon_\ast = \big[\Delta F(T) -T d \Delta F(T)/dT
\big]\Big|_{T=T_{\ast}} $ over radiation energy density where $T_\ast$ and
$\Delta F(T)$ are the temperature of the thermal bath at time of production
of GWs and the difference of the free energy between two phases respectively, and
$S_{\textup{en}}, S_{\textup{sw}}$, and $ S_{\textup{tu}} $ are spectral
shapes of GWs which are characterized from numerical fits
as~\cite{Huber:2008hg, Caprini:2015zlo, Hindmarsh:2015qta, Caprini:2009yp} 
\bea
S_{\textup{en}} &=& \frac{3.8 \left(\frac{f}{f_{\textup{en}}} \right)^{2.8}}{1+2.8 \left(\frac{f}{f_{\textup{en}}}\right)^{3.8}} \,, \nn \\
S_{\textup{sw}} &=& \left(\frac{f}{f_{\textup{sw}}} \right)^3 \left(\frac{7}{4 +3 \left(\frac{f}{f_{\textup{sw}}}\right)^{2}} \right)^\frac{7}{2} \,, \nn \\
S_{\textup{tu}}  &=& \frac{\left(\frac{f}{f_{\textup{tu}}} \right)^3 }{\left(1+ \frac{f}{f_{\textup{tu}}} \right)^{\frac{11}{3}} \left(1 +\frac{ 8 \pi f}{h_{\ast}}\right)} \,,
\eea
where $h_{\ast}$ is Hubble rate at $T_\ast$, $f_{\textup{en}},
f_{\textup{sw}}$ and $f_{\textup{tu}}$ are peak frequencies in three cases,
which are given by~\cite{Kamionkowski:1993fg, Caprini:2015zlo,
Caprini:2009yp}
\bea
h_{\ast} &=& 11.2\ \textup{nHz}\ \left(\frac{T_\ast}{100 \textup{MeV}} \right) \left(\frac{g_\ast}{10}\right)^{\frac16} \,, \nn \\
f_{\textup{en}} &=& 11.2\ \textup{nHz}\ \left(\frac{0.62}{1.8 -0.1 v_w +v_w^2}\right) \left(\frac{\beta}{H_\ast}\right) \left(\frac{T_\ast}{100 \textup{MeV}}\right) \left(\frac{g_\ast}{10} \right)^{\frac16} \,, \nn \\
f_{\textup{sw}} &=& 12.9 \ \textup{nHz}\ \left(\frac{1}{v_w}\right) \left(\frac{\beta}{H_\ast}\right) \left(\frac{T_\ast}{100 \textup{MeV}}\right) \left(\frac{g_\ast}{10} \right)^{\frac16} \,,\nn \\
f_{\textup{tu}} &=& 18.4 \ \textup{nHz}\ \left(\frac{1}{v_w}\right) \left(\frac{\beta}{H_\ast}\right) \left(\frac{T_\ast}{100 \textup{MeV}}\right) \left(\frac{g_\ast}{10} \right)^{\frac16} \,.
\eea
 In Eq.~(\ref{eq:Omega}), $\kappa_1 , \kappa_2 $, and $ \kappa_3 $ are the
 fractions of the vacuum energy converted to the kinetic energy of the
 bubbles, bulk fluid motion, and the MHD turbulence, respectively. These
 factors are model-dependent. In this work, we consider two cases of
 bubble: Jouguet detonations and non-runaway bubbles. For the case of
 Jouguet detonations~\cite{Steinhardt:1981ct, Caprini:2015zlo,
 Kamionkowski:1993fg, Nicolis:2003tg, Espinosa:2010hh, Hindmarsh:2015qta},
 we have
 \bea
 \kappa_1 &=& \frac{0.715 \alpha +0.181 \sqrt{\alpha}}{1+ 0.715 \alpha}  \,,\nn \\
 \kappa_2 &=& \frac{\sqrt{\alpha}}{ 0.135 + \sqrt{\alpha +0.98} } \,, \nn \\
 \kappa_3 &=& 0.05 \kappa_2 \,, \nn \\
 v_w &=& \frac{\sqrt{1/3} +\sqrt{\alpha^2 +2 \alpha/3} }{ 1+ \alpha } \,,
 \eea
and for the case of non-runaway bubbles~\cite{Espinosa:2010hh, Caprini:2015zlo,
Hindmarsh:2015qta, Kamionkowski:1993fg},
\bea
\kappa_1 &=& 0 \,, \nn \\
\kappa_2 &=& \frac{\alpha}{0.73 +0.083 \sqrt{\alpha} +\alpha } \,, \nn \\
\kappa_3 &=& 0.05 \kappa_2 \,, \nn \\
v_w &=& 0.95 \,.
\eea
With all these expressions, there are still four
unknown parameters in Eq.~(\ref{GW}), i.e., $g_\ast, \beta/H_\ast$, $T_\ast$ and $\alpha$,  which characterize
the first-order cosmological QCD phase transition. Generally, for QCD phase
transitions, the temperature $T_\ast$ is around several hundreds MeV, of which the concrete value depends on the types of the QCD matter and the phase transition.
We will obtain, in the following section,  $T_\ast$  for the phase transition in different QCD
matters by the method of the holographic QCD. One can also calculate  $\alpha$ from different holographic models. For simplicity, we assume that the GW can be generated soon after the confinement/deconfinement phase transition occurs, so $T_\ast$ is approximated by the critical temperature of the phase transition. Besides, we fix
$g_\ast$ and  $\beta/H_\ast $ at their typical values which could be chosen as $g_\ast =10$ and $\beta/H_\ast =10$~\cite{Ahmadvand:2017xrw, Ahmadvand:2017tue,Chen:2017cyc} at the QCD scale. 

\section{GWs from holographic QCD models and the NANOGrav signal}

In this section, we will obtain the GW spectra  from the first-order confinement/deconfinement transition in heavy static
quarks with a zero baryon chemical potential, quarks with a finite baryon
chemical potential, and a pure gluon system, and match them with the
NANOGrav signal. First, we start with finding the corresponding
critical temperatures by use of holographic QCD models.

Following the Anti-de Sitter/conformal field theory (AdS/CFT)
correspondence principle~\cite{Maldacena:1997re, Gubser:1998bc,
Witten:1998qj}, AdS/QCD offers some new insights to the non-perturbative
hadron dynamics from the dual gravitational field~\cite{Erlich:2005qh}.
Especially, the first-order confinement/deconfinement phase transitions
could be interpreted by Hawking-Page (HP) phase
transitions~\cite{Hawking:1982dh} in five-dimensional spacetime in the
AdS/QCD models~\cite{Herzog:2006ra}, where the high-temperature QGP
corresponds to the AdS black hole, while the low-temperature hadron corresponds
to the thermal AdS space. In the absence of the baryon chemical potential, the
transition temperature calculated by the soft-wall model is consistent with
numerical results~\cite{Herzog:2006ra}. In the case of the finite baryon
chemical potential, AdS/QCD models can also give a good explanation of the phase transition~\cite{Horigome:2006xu, Kim:2007em,
Seo:2009kg, Cai:2012xh}, while the
standard lattice QCD simulation suffers from the famous sign problem~\cite{Takeda:2011vd,
Fromm:2011qi, Ding:2017giu} and could not provide many useful results. The
GW produced by the first-order QCD phase transition was estimated in the case
of the heavy static quark system with a zero baryon chemical potential via
hard-wall and soft-wall models of AdS/QCD in Ref.~\cite{Ahmadvand:2017xrw}
for the first time, and then studied in the finite chemical potential
system~\cite{Ahmadvand:2017tue, Ahmadvand:2020fqv} and pure gluon
system~\cite{Chen:2017cyc} via different models. It is also worth
mentioning that the initial idea of explaining GWs generated from 
cosmological phase transitions by holographic method could be traced back to
Randall and Servant's seminal work~\cite{Randall:2006py} in 2006 to the
best of our knowledge.

We consider three types of QCD matter systems in five different holographic
models: (i) heavy static quarks with a zero baryon chemical potential in
hard-wall model $S_1$ and (ii) soft-wall model $S_2$, (iii) quarks with a
finite baryon chemical potential in hard-wall model $S_3$ and (iv)
soft-wall model $S_4$, and (v) pure gluons in the quenched dynamical
holographic model $S_5$. The corresponding five-dimensional gravitational
actions are given by~\cite{Herzog:2006ra, Ahmadvand:2017tue,
Ahmadvand:2017xrw, Chen:2017cyc}
\bea
S_1 &=& \frac{1}{2\kappa^2}  \int d^5 x \sqrt{-g}  \left(R +\frac{12}{\ell^2} \right) \,, \\
S_2 &=& \frac{1}{2\kappa^2} \int d^5 x \sqrt{-g}  e^{-\phi} \left(R +\frac{12}{\ell^2} \right) \,, \\
S_3 &=& \int d^5 x \sqrt{-g}  \left[\frac{1}{2\kappa^2}  \left(R +\frac{12}{\ell^2} \right) -\frac{1}{4 g_5^2} F^2 \right] -\frac{1}{\kappa^2} \int d^4 x \sqrt{h}  \nabla_\mu  n^\mu   \,, \\
S_4 &=& \frac{1}{2\kappa^2} \int d^5 x \sqrt{-g} e^{-\phi} \left[\frac{1}{2\kappa^2}  \left(R +\frac{12}{\ell^2} \right) -\frac{1}{4 g_5^2} F^2 \right] -\frac{1}{\kappa^2} \int d^4 x \sqrt{h} e^{-\phi} \nabla_\mu  n^\mu  \,, \\
S_5 &=& \frac{1}{2\kappa^2} \int d^5 x \sqrt{-g} e^{-2 \varphi} \left(R -\frac{4}{3}\partial_\mu \varphi  \partial^\mu \varphi - V(\varphi) \right) \,, 
\eea
where $\kappa^2 = 8 \pi G$, $g_5$ is coupling constant, $\ell$ is the radius of five-dimensional AdS space, $g$ and $h$ are determinants of bulk and boundary metrics respectively, $n^\mu $ is the unit vector normal to the hypersurface, $\phi $ is a non-dynamical dilaton,  and $\varphi $ is a dynamical dilaton. The confinement/deconfinement transition in the cases of heavy static quarks with a zero baryon chemical potential and pure gluons are analogous to the HP transition between the static AdS black hole and the thermal AdS vacuum. For the  case of quarks with a finite baryon chemical potential, the confinement/deconfinement transition is analogous to the HP transition between the charged AdS black hole and the thermal charged AdS vacuum, and the chemical potential is related to the electric charge of  the black hole. For a specific model, one can calculate the free energies of the black hole and AdS vacuum respectively, and obtain  the difference of the free energy $\Delta F$ between two phases. Then one can obtain the value of the temperature $T_\ast$ and $\alpha$ via some holographic techniques. Here, we refer readers to Refs.~\cite{Herzog:2006ra, Ahmadvand:2017tue, Ahmadvand:2017xrw, Chen:2017cyc} for detailed calculations and  list the corresponding critical temperatures in these models in Table \ref{tab1}.

\begin{table}[htbp]
  \caption{Critical temperatures from five holographical QCD models.}
  \begin{center}
  \begin{tabular}{ lll }
  \hline\hline
  QCD matters & Holographic QCD models & Temperature  \\ 
   \hline
  {heavy static quarks with} & hard wall  & 122 MeV ~\cite{Herzog:2006ra, Ahmadvand:2017xrw}\\ 
   
   a zero chemical potential & soft wall & 191 MeV~\cite{Herzog:2006ra, Ahmadvand:2017xrw} \\ 
   \hline
  quarks with a finite & hard wall & 112 MeV~\cite{Ahmadvand:2017tue}\\ 
   
   chemical potential & soft wall  & 192 MeV~\cite{Ahmadvand:2017tue} \\ 
   \hline
  pure gluons & quenched dynamical holographic QCD & 255 MeV~\cite{Chen:2017cyc} \\ 
   \hline
  \end{tabular}  
  \label{tab1}
  \end{center}
  \end{table}

\begin{figure}[htbp]
\centerline{
 \includegraphics[width=0.95\linewidth]{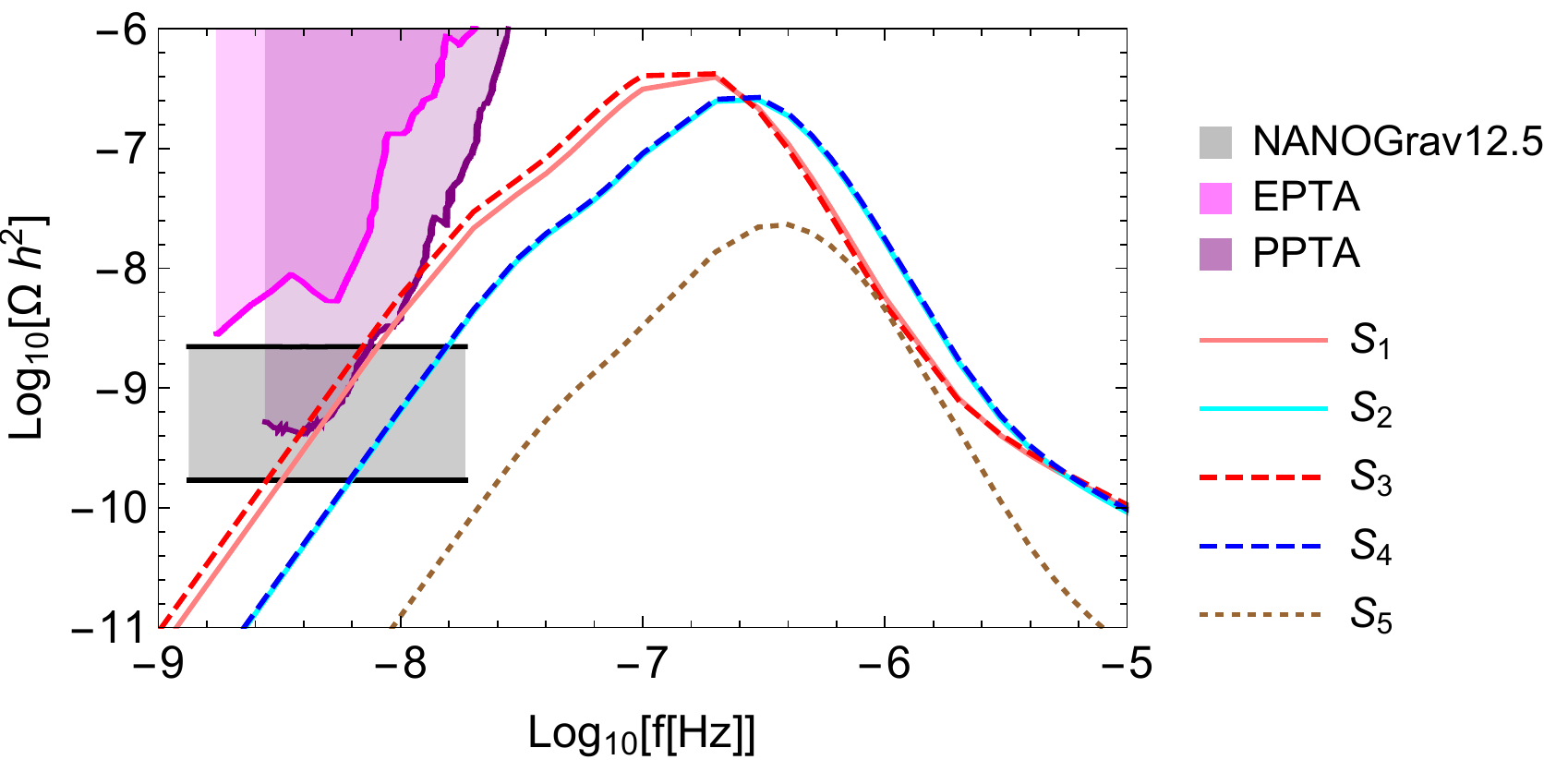}  \
 } 
 \caption{GW spectra from QCD matter confinement/deconfinement phase
 transition in the Jouguet detonation bubble case from five holographical
 models. The results are compared with the current sensitivity of
 NANOGrav 12.5-yr~\cite{Arzoumanian:2020vkk, DeLuca:2020agl}, EPTA~\cite{Desvignes:2016yex}, and
 PPTA~\cite{Kerr:2020qdo}.}
 \label{figjouguet}
\end{figure}

\begin{figure}[htbp]
\centerline{
\includegraphics[width=0.95\linewidth]{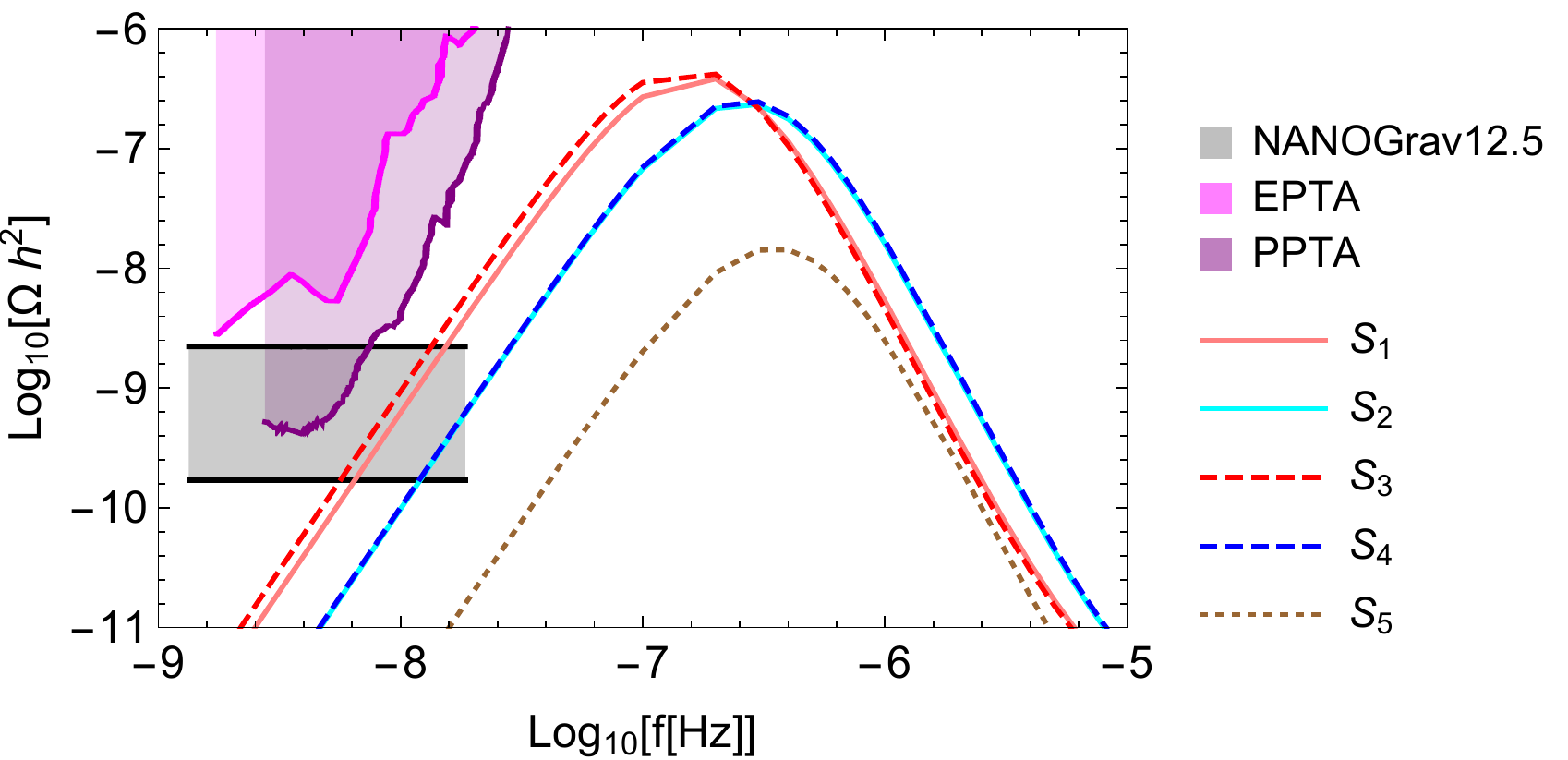} \
 }
 \caption{Same as Fig.~\ref{figjouguet}, for the non-runaway bubble case.}
 \label{fignonrunaway}
\end{figure}

We assume that GW is generated quickly after the phase transition occurs and 
temperature $T_\ast$ is approximated as the critical phase transition
temperature. Now we match the NANOGrav results with the GW produced from
the confinement/deconfinement phase transition in different holographic
models. We plot the GW spectra in two bubble models: Jouguet detonations
and non-runaway bubbles. The results are illustrated respectively in
Fig.~\ref{figjouguet} and Fig.~\ref{fignonrunaway}.

 From Fig.~\ref{figjouguet} and Fig.~\ref{fignonrunaway}, we show that the power spectra of GWs 
 from the quark confinement phase transitions enter  the 95\% confidence interval from 
 the NANOGrav 12.5-yr observation~\cite{Arzoumanian:2020vkk, DeLuca:2020agl}, which indicates that 
 the confinement/deconfinement phase transition in pure quark systems (cases of heavy static quarks with a zero baryon chemical potential and quarks with a finite baryon chemical potential) could 
 possibly be the potential cosmological sources of the NANOGrav signal for both the Jouguet
 detonation and non-runaway bubble cases. 
 The power spectra of GWs calculated from different holographic models (hard wall and soft wall) in a specific QCD matter system are different, but the difference  does not change the conclusion.  In contrast, the confinement/deconfinement transition in the pure gluon system could not possibly be the source of the NANOGrav signal. Since the critical phase transition temperatures for the cases of a finite chemical potential and a zero chemical potential are very close, the chemical potential has little effect on
 the power spectrum of GWs, and the quark confinement dominates the
 cosmological QCD transition according to the current NANOGrav observation. But
 these conclusions need to be further supported by more accurate
 observations of GWs in the future. We also calculate the spectrum indices
 from different holographic models, and find that the values are similar in the
 same bubble model. The values are about 2.78 and 2.91 for  the Jouguet and
 non-runaway cases, respectively. These values could be used to check our conclusions with more accurate  future observations.

\section{Conclusions and Discussions }

In this paper, we have showed that a possible stochastic
common-spectrum process reported by the NANOGrav Collaboration based on the
12.5-yr PTA data can be explained potentially as a GW signal from the
first-order cosmological confinement/deconfinement phase transition in the
cases of (i) heavy static quarks with a zero baryon chemical potential and
(ii) quarks with a finite baryon chemical potential. We also find that the gluon confinement could not possibly be the potential source of the NANOGrav signal based on the current observation. We match the GW spectra  from the first-order phase transition in five different holographic QCD models
with the NANOGrav signal. By considering  both the Jouguet
detonation and non-runaway bubble growth models, we find that the GW spectra from the confinement/deconfinement phase transition in pure quark systems, irrespective of whether  the  baryon chemical potential is finite or zero, enter the 95\% confidence interval from the NANOGrav 12.5-yr observation~\cite{Arzoumanian:2020vkk, DeLuca:2020agl}.
 The baryon chemical potential has little influence on the power
spectra of GWs produced by confinement/deconfinement phase transitions since the phase transition temperatures in both the finite baryon chemical potential case and zero chemical potential case are very close. 

We must point out that we have assumed in our analysis that GWs are generated soon after the phase transition happens,
so the temperature at which the GWs are produced is approximately the
transition temperature.  Although this is an acceptable assumption,  it remains interesting 
to find a more credible holographic method to calculate the
temperature $T_\ast$, analogous to what was done in
Ref.~\cite{Randall:2006py}. Moreover,  it
is also worth considering the confinement/deconfinement phase transitions
in  possible QCD matter systems other than those  we have looked at in the present paper  and chiral symmetry breaking phase transitions. We would like to 
leave these to future studies.

\section*{Acknowledgement }

We are grateful to Jie-Wen Chen, Muyang Chen, Chengjie Fu, Yong Gao, Long-Cheng Gui, Hongbo Li, Chang Liu, Jing Liu, H. L\"u, Xueli Miao, Shi Pi, Shao-Jiang Wang, Hao Wei, Rui Xu, and Junjie Zhao for useful discussions. SL thanks Lijing for his warm hospitality during the visit to KIAA. SL, PW and HY were supported in part by the NSFC under Grants No. 11947216, No. 11690034, No. 11805063,  No. 11775077  and No. 12075084,  and China Postdoctoral Science Foundation 2019M662785.
LS was supported by the National SKA Program of China (2020SKA0120300), the
National Natural Science Foundation of China (11975027, 11991053,
11721303), the Young Elite Scientists Sponsorship Program by the China
Association for Science and Technology (2018QNRC001), and the Max Planck
Partner Group Program funded by the Max Planck Society.

\end{document}